# Regulating electron diffraction direction with cylindrically symmetric rotating crystal

L. Cheng[1], B. Da[2,3*], X. Liu[1,2], K. Shigeto[4], K. Tsukagoshi[5], T. Nabatame[5], J. W. Liu[6], H. Zhang[3], H. Yoshikawa[2,3], and S. Tanuma[3], Z. S. Gao[7], H. X. Guo[8], Y. Sun[9], J. Hu[10], Z. J. Ding[1#]

[1]*Department of Physics, University of Science and Technology of China, Hefei, Auhui 230026, China*

[2]*Research and Services Division of Materials Data and Integrated Systems, National Institute for Materials Science, 1-1 Namiki, Tsukuba, Ibaraki 305-0044, Japan*

[3]*Research Center for Advanced Measurement and Characterization, National Institute for Materials Science, 1-1 Namiki, Tsukuba, Ibaraki 305-0044, Japan*

[4]*Hitachi High-Tech Corporation, Hitachinaka, Ibaraki 312-8504, Japan*

[5]*International Center for Materials Nanoarchitectonics (WPI-MANA), National Institute for Materials Science, Tsukuba, Ibaraki 305-0044, Japan*

[6]*Research Center for Functional Materials, National Institute for Materials Science, 1-1 Namiki, Tsukuba, Ibaraki 305-0044, Japan*

[7]*Institute of Electrical Engineering, Chinese Academy of Sciences, Beijing, 100190, P.R. China.*

[8]*Key Laboratory of MEMS of Ministry of Education, School of Electronic Science and Engineering, Southeast University, Nanjing, 210096, People's Republic of China*

[9]*Department of Applied Physics and Applied Mathematics, Columbia University, New York, NY, 10027, USA*

[10]*Department of Physics and Institute for Nanoscience and Engineering, University of Arkansas, Fayetteville, Arkansas 72701, United States*

*DA.Bo@nims.go.jp

#zjding@ustc.edu.cn

## ABSTRACT

We report a promising InSiO film that allows simultaneous observation of sample morphology and Kikuchi patterns in raster scan mode of scanning electron microscopy. This new experimental observation suggests potential mechanism beyond existing diffraction theories. We find by simulation that this material has a novel cylindrically symmetric rotational crystalline structure that can control the diffraction direction of electrons through a specific rotational distribution of crystal planes, while being independent of the angle and energy of incident electrons within a certain range.

## MAIN TEXT

Electron microscope is an outstanding tool for material characterization [1,2]. Various electron microscopy techniques have been developed by analyzing signals from electron-material interactions. In general, these techniques can be divided into several categories: Energy spectroscopy, such as reflection electron energy loss spectroscopy [3-11], secondary electron spectroscopy [12-16], and Auger electron spectroscopy [17-20]; Electron diffraction patterns, for example, selected-area electron diffraction (SAED) [21], low-energy electron diffraction [22], and precession electron diffraction [23]; Electron microscopic imaging, as for backscattered electron (BSE) [24-27], secondary electron imaging [28-31], high-angle annular dark field [32]. These well-established techniques are based on different physical processes. Energy spectroscopy mainly analyzes the scattering processes of electrons [33-37] and is often used to study the electron structure of the sample. Electron diffraction pattern focuses on the diffraction process of incident electrons in the lattice [30,38-41] and is commonly used to study the crystal structure. Electron microscopic imaging concentrates on the amplitude or phase contrast of electrons [24,30,42,43], and is usually used to observe the morphology and structure of materials.

Electron microscopes are conventionally designed with many modes for analyzing different signals. In scanning electron microscopy (SEM), there is a consensus to use the raster scan mode to observe the surface morphology and the grazing incidence mode to observe the electron backscatter diffraction (EBSD) pattern [44,45]. The reason is that in the grazing incidence mode, the incident electrons undergo a large amount of inelastic scattering and each atom becomes a point source allowing the electrons to be incident into the lattice in all directions, equivalent to spherical subwave sources [40,41,46]. Similarly, electron channeling diffraction (ECP) [47] uses a rocking beam to directly incident electrons into lattices from different directions. Either way, it is generally recognized by existing experimental techniques and physical theories that electrons entering the lattice successively/simultaneously at continuously varying angles is necessary to produce the Kikuchi pattern.

However, we have found that InSiO film exhibits both the island morphology and crystal Kikuchi pattern in the raster scan mode of SEM without introducing spherical wave or rocking beam. This phenomenon, never reported previously, differs from the usual experimental phenomena, and it could even theoretically break our knowledge about the Kikuchi pattern formation mechanism. The physical processes involved are worth exploring.

In this letter, we investigate the physical mechanism of the interaction of electrons with this InSiO material. A dynamical simulation approach is developed. First, BSE images of InSiO will be shown to illustrate why the patterns are considered to be Kikuchi bands in momentum-space, rather than nanowire structures in real-space. Then we propose a potential physical mechanism for the formation of the Kikuchi band that does not require spherical wave or rocking beam, but rather a specific structure of the sample. By comparing experiments with simulations, we determined that the presented InSiO has a cylindrically symmetric rotating crystal (CSRC) structure. This special CSRC structure is able to modulate the diffraction of electrons and thus allowing the superposition of the diffraction signal in momentum-space with the surface morphology signal in real-space. In addition, the CSRC structure gives a new idea of manipulating electrons, which enables the control of the scattering angle of electrons at the micro-nano scale. We suggest that this particular structure has crucial implications for the development of physical science and device components.

The dynamical methodology describes the wave function inside the crystal as a superposition of Bloch waves [48-50],

$$\psi(\mathbf{r}) = \sum_j A^{(j)} \sum_{\mathbf{g}} C_{\mathbf{g}}^{(j)} \exp\left[2\pi i \left(\mathbf{k}^{(j)} + \mathbf{g}\right) \cdot \mathbf{r}\right], \quad (1)$$

where $A^{(j)}$ and $C_{\mathbf{g}}^{(j)}$ are coefficients, $\mathbf{g}$ is the reciprocal vector, and $\mathbf{k}^{(j)}$ is the wave vector of the $j$th branch of Bloch wave. The wave function $\psi(\mathbf{r})$ is a solution of the Schrödinger equation. Substituting the wave function into the Schrödinger equation yields Bethe's dynamical equation [51],

$$\left[\mathbf{K}^2 - \left(\mathbf{k}^{(j)} + \mathbf{g}\right)^2\right] C_{\mathbf{g}}^{(j)} + \sum_{\mathbf{h} \neq \mathbf{g}} U_{\mathbf{g}-\mathbf{h}} C_{\mathbf{h}}^{(j)} = 0, \quad (2)$$

where $\mathbf{K}^2 = \mathbf{k}_0^2 + U_0$, $\mathbf{K}$ and $\mathbf{k}_0$ are the wave vectors of incident electrons inside the crystal and in the vacuum, respectively. $U_\mathbf{g} = 2meV_\mathbf{g}/h^2$ is the electron structure factor, where *m* is the relativistic electron mass, *e* is the elementary charge, *h* is the Planck constant, and $V_\mathbf{g}$ is the Fourier coefficient of the crystal potential, $V(\mathbf{r}) = \sum_\mathbf{g} V_\mathbf{g} \exp(2\pi i \mathbf{g} \cdot \mathbf{r})$. In general, crystal potential is extended to complex numbers to introduce inelastic scattering processes [52,53].

The Bloch wave vector $\mathbf{k}^{(j)}$ is decomposed as $\mathbf{k}^{(j)} = \mathbf{K} + \lambda^{(j)}\hat{\mathbf{n}}$, where $\hat{\mathbf{n}}$ is the unit normal vector of the sample surface. Then the Bethe's equation (2) is transformed into an eigenvalue problem [54] with complex $\lambda^{(j)}$ as the eigenvalues and $\mathbf{C}^{(j)}$ as the the eigenvectors whose elements are $C_\mathbf{g}^{(j)}$. This eigenvalue problem can be solved by stable methods [55,56] and the Bloch wave excitation coefficients $A^{(j)} = C_j^{-1}$ can be derived from the boundary conditions on the crystal surface [48,57]. Here $C_j^{-1}$ is the *j*th element of the first column of $\mathbf{C}^{-1}$, which is the inverse matrix of **C**, where **C** is aligned by the eigenvectors $\mathbf{C}^{(j)}$. After solving all required parameters, Eq. (2) can be rewritten as,

$$\psi(\mathbf{r}) = \sum_j A^{(j)} \sum_\mathbf{g} C_\mathbf{g}^{(j)} \exp\left[2\pi i \left(\mathbf{K} + \lambda^{(j)}\hat{\mathbf{n}} + \mathbf{g}\right) \cdot \mathbf{r}\right]. \quad (3)$$

The probability density $P(\mathbf{r})$ of any position inside the crystal is as follows,

$$P(\mathbf{r}) = \psi\psi^* = \sum_{i,j} \sum_{\mathbf{g},\mathbf{h}} A^{(i)} A^{(j)*} C_\mathbf{g}^{(i)} C_\mathbf{h}^{(j)*} \exp\left[2\pi i \left(\gamma^{(i)} - \gamma^{(j)*}\right)\hat{\mathbf{n}} \cdot \mathbf{r}\right] \exp[2\pi i (\mathbf{g}-\mathbf{h}) \cdot \mathbf{r}] \quad (4)$$

It is necessary to note that $P(\mathbf{r})$ is **K**-dependent even though there is no **K** in Eq. (4). When **K** changes, all solved coefficients also varies, obtaining various wave functions in Eq. (3), and therefore the probability density $P(\mathbf{r})$ is different. In other words, $P(\mathbf{r})$, or $P(\mathbf{K},\mathbf{r})$, represents the diffraction intensity concerning an observed direction **K** or a crystal orientation [*hkl*]. For any direction **K**, the observed intensity $I(\mathbf{K})$ is calculated by integrating all points in the crystal,

$$I(\mathbf{K}) \propto \int P(\mathbf{K},\mathbf{r})d\mathbf{r}. \quad (5)$$

In raster scan mode, the wave vector **K** of the electron beam is invariant. For a general crystal, the relative direction of the wave vector to an arbitrary crystal orientation [*hkl*] is also constant. However, since the crystal orientation of CSRC is different everywhere, as in Fig. 1, the direction of **K** cannot be defined by the ground but by the crystal orientation [*hkl*] at the incident position as the reference system, i.e., **K** is also different everywhere. For any given incident position, the crystal direction has to be identified before defining the wave vector of electrons. Then the diffraction intensity at that position can be obtained by Eqs. (3-5). The diffraction pattern of the whole CSRC island can be derived by calculating and reconstructing the diffraction intensity at all positions according to the crystal direction distribution.

Amorphous films with a thickness of 30 nm were prepared on a sapphire substrate at room temperature by DC magnetron sputtering. The sputtering targets consisted of $In_2O_3$ and $SiO_2$ with a Si/In ratio of 2.3 at.%, equivalent to 1 wt.% of $SiO_2/(In_2O_3+SiO_2)$. The amorphous films were then annealed at 300 °C and gradually crystallized into CSRC islands with diameters of about 1-2 μm. Detailed information could be found in Ref. [58].

SEM image of the InSiO film is displayed in Fig 2(a). It can be observed that there are many round islands with some band patterns inside. Generally, Fig. 2(a) is understood as some crystal islands with nanowire structure inside. However, when we select the crystal island in the red rectangular region of Fig. 2(a) for further observation, we find that when the sample is tilted, as in Fig. 2(b), the band patterns are not fixed but relatively shifted on the island. This anomalous displacement indicates that the band patterns cannot be nanowire structures of the crystal island but originate from some other physical processes. Meanwhile, when the incident electron energy increases, as shown in Fig. 2(c), the shape of the band pattern is not constant but gradually narrower. Such features suggest that the band patterns may originate from electron diffraction.

Common electron diffraction patterns, such as SAED, are dot arrays whose diffraction angle is mainly determined by incident electron direction. When the crystal is tilted/rotated, the direction of the diffraction patterns will not change obviously, but the distribution will vary drastically depending on the crystal orientation. In contrast, Kikuchi patterns are band patterns and appear to be fixed on the crystal. When tilting/rotating the crystal, the diffraction pattern will rotate together with the sample. Therefore, the band patterns in CSRC are probably Kikuchi patterns.

However, according to general experimental experience and existing electron diffraction theory, the raster scan mode is not expected to produce Kikuchi patterns. The contrast of the Kikuchi band originates from the coherent superposition of electron diffraction wave functions. When the electron sources have a single direction, the diffraction patterns are in the form of dot arrays. While the electron source rocks within a small cone angle, the dot arrays expand into discs. As the cone angle increases further, the discs intersect and overlap, thus coherently superposing and forming Kikuchi patterns. The existing experimental techniques obtain the Kikuchi pattern by different means. For example, the ECP and EBSD use rocking beam and grazing incidence (equivalent spherical subwave) to directly/indirectly increase the range of electron incidence angles, respectively. Nevertheless, one cannot obtain the Kikuchi pattern directly from the raster scan mode because its electron source has a single direction. Hence, we believe that the unknown anomalous mechanisms generating the Kikuchi patterns come from some nature of the CSRC itself.

Here we propose a reasonable hypothesis for the crystal structure of CSRC. Notice that in ECP experiments, the rocking beam can be considered as many electron beams incident into the crystal in different directions. For any of these beams $b_i$, there is a definite incidence position $r_i$ and an incidence angle $\theta_i$, precisely, the polar angle $\theta_i$ and the azimuthal angle $\varphi_i$. The incidence angle is usually defined as the angle between the incident electron beam and the surface normal of the sample. But here we redefine it as the angle between the electron beam and the crystal direction [*hkl*], and for general crystals, the two definitions are equivalent due to translational symmetry. Then for each scanned position $r_i$ in the raster scan mode (see Fig. 1), if the crystal direction is rotated to the same angle as the corresponding angle $\theta_i$ and $\varphi_i$ in ECP mode, this results in a special crystal, the CSRC, with broken translational symmetry but with extra rotational symmetry. Ideally, performing raster scan experiments with this CSRC is equivalent to an ECP experiment with a conventional crystal.

Based on the above hypothesis, we build an ideal CSRC island. First, an arbitrary crystal orientation, [*hkl*], is selected and set at the center of the crystal island. Then the crystal orientation at each surrounding point is rotated once, and the azimuthal and polar angles of the rotated crystal orientation with respect to the original orientation are described in Fig. 3(a). In our previous research [59], XRD experiments showed that only diffraction peaks of bixbyite $In_2O_3$ and no peaks of $SiO_2$ were observed in the presented InSiO CSRC sample, which proves that the Si atoms are soluble in the $In_2O_3$ matrix and the diffraction intensity is mainly contributed by $In_2O_3$. Therefore, it can be assumed that each point in Fig. 3(a) corresponds to a bixbyite type $In_2O_3$ grain with a different crystal orientation. Then the Kikuchi pattern can be obtained by the present dynamical method.

Figure 3(b) shows a simulated Kikuchi pattern of an ideal CSRC island with incident electron energy of 15 keV and crystal direction of [440] at the center position, as well as an experimental BSE image. The characteristic Kikuchi patterns are clearly observed in the experimental and simulated figures, which are marked as paired red lines. There are six major Kikuchi bands intersecting at one pole. The Kikuchi bands originate from the Bragg reflections of sets of crystal planes, whose indices are marked in red brackets, while the pole represents the [440] zone axis. The simulated Kikuchi patterns agree well with the experimentally observed ones in the presented BSE images, indicating that our proposed hypothesis appropriately explains the mechanism of the generation of the Kikuchi band when the raster scanning electron beam diffracted in CSRC. Next, we further validate our theoretical model by EBSD experiments. Figure 3(c) shows the deviation of the grain reference orientation of the CSRC island in Fig. 3(b), where the deviations of the grain axis and angle are exhibited in the left and right panels, respectively. What they depict actually corresponds to the crystal plane/orientation rotation described by azimuthal and polar angles in Fig. 3(a). It is evident that Fig. 3(c) is very similar to Fig. 3(a) in that the crystallographic axes/azimuths rotate a full circle, while the angular/polar angles rotate uniformly and slowly from the center to the periphery. In other words, the experimentally observed rotation distribution of the crystal plane/orientation is the same as our hypothesis.

With this method, arbitrary CSRC islands can be calculated. We simulate each crystal island according to the directions marked in Fig. 2(a) and combine them in Fig. 4(a). The number, shape, and bandwidth of the simulated Kikuchi bands are consistent with the experimental results, which proves the validity of our theoretical model. We also simulate different sample tilt angles and electron incident energies in Figs. 4(b) and (c) for comparison with Figs. 2(b) and (c), and similarly observed the shift of the Kikuchi pattern as well as the variation of the bandwidth, respectively. Theoretically, the Kikuchi pattern will rotate with the crystal orientation, so the displacement of the Kikuchi pattern will be proportional to the tilt angle of the sample at a small-angle approximation. Meanwhile, the angular width of the Kikuchi band is equal to twice the corresponding Bragg angle. According to Bragg's law, $\lambda = 2d \sin \theta_B$, where $\lambda$ is the wavelength of the incident electrons, $d$ is the lattice plane spacing, and $\theta_B$ is the Bragg angle. For small angle cases, $\sin \theta_B \approx \theta_B$, thus the width of the Kikuchi band is in proportion to the wavelength of the incident electrons. As for the CSRC, it shows in Fig. 4(d) that the shift distance is proportional to the tilt angle of the sample in the range of $\pm 5°$. Similarly, Fig. 4(e) also presents the width of a selected band from Fig. 2(c) and Fig. 4(c) as a function of the wavelength of the incident electrons in a proportional manner.

Naturally, there are also some differences in details between the experimental and simulated results. It is because the experimental images also contain the surface morphology information obtained by raster scanning, such as the white spots in the experimental images in Fig. 3(b), which are irregular particles rather than diffraction patterns of the sample when it crystallizes. On the other hand, the simulated ideal CSRC island has a relatively consistent rotation velocity of the crystal plane, while the experimental specimen may have a local anomalous distortion, which can lead to blurring of the experimental Kikuchi band.

We have simultaneously observed the surface morphology and the Kikuchi pattern in a unique sample, InSiO CSRC by SEM in the raster scan mode and developed a theoretical model to study this phenomenon. The formation mechanism of the Kikuchi pattern is different from the conventional theory. CSRC achieves an electron diffraction environment equivalent to ECP through a particular rotational distribution of the crystal plane. It is important to note that within an appropriate angular range, the distribution of the Kikuchi pattern is principally related to the structure of the CSRC and is independent of the direction of the incident electron beam. In addition, the energy of the incident electrons only affects the Kikuchi bandwidth, while the angle of the electron diffraction center is constant. In other words, CSRC can regulate the emission direction of electrons only by the rotation distribution of the crystal plane regardless of the electron energy, which gives a new idea of electron manipulation.

This work was supported by the National Institute for Materials Science under the Support system for curiosity-driven research, JSPS KAKENHI Grant Number JP21K14656, Grant for Basic Science Research Projects from The Sumitomo Foundation and from The Kao Foundation for Arts and Sciences. J.H. was supported by the U.S. Department of Energy (DOE), Office of Science, Office of Basic Energy Sciences under Award DE-SC0019467. Z.J.D was supported by the National Key Research and Development Project (2019YFF0216404) and Education Ministry through "111 Project 2.0" (BP0719016). Calculations were performed on the supercomputing center of USTC and the Numerical Materials Simulator supercomputer at the National Institute for Materials Science. We are thankful for helpful discussions and suggestions from Dr. Takio Kizu.

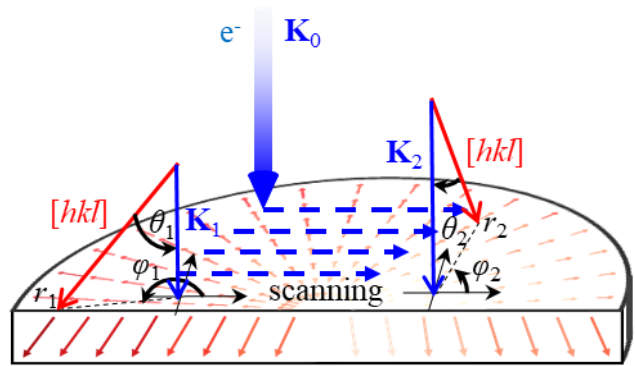

FIG. 1. Schematic diagram of the structure of CSRC and the dynamical method to calculate the diffraction intensity of electrons incident at any position. The red and blue arrows represent the crystal direction [*hkl*] and the electron wave vector **K** at the electron incidence position, respectively.

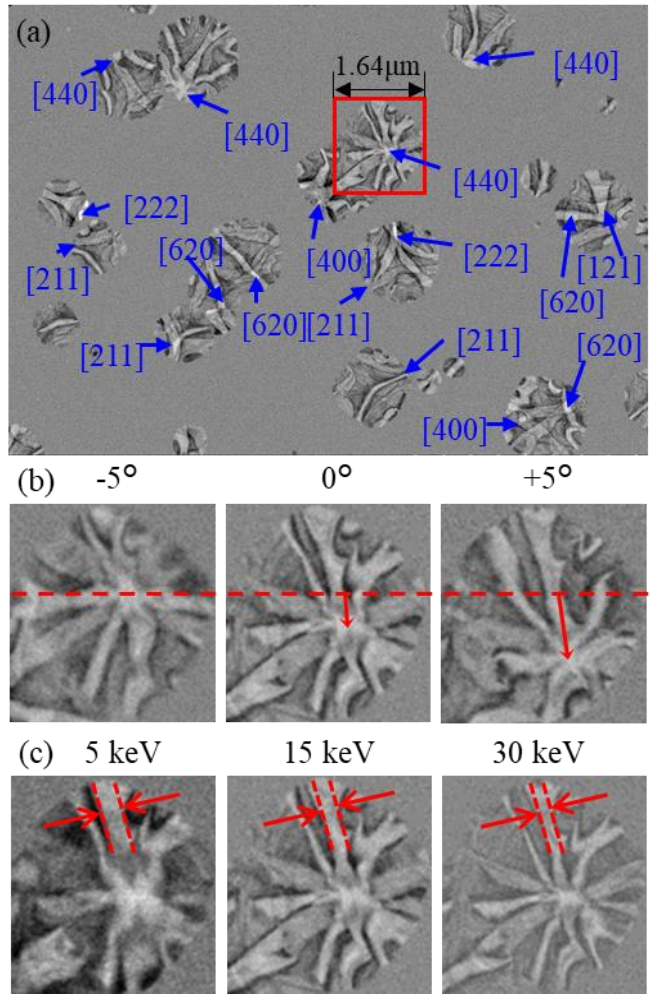

FIG. 2. SEM images of InSiO CSRC islands obtained by BSE detector. (a) Image of multiple islands at an incident electron energy of 15 keV during the crystallization process. The indices in blue brackets represent the direction of the crystal at the position indicated by the blue arrows. (b, c) BSE images measured from the [440] crystal island in the red rectangular region in (a) at different (b) tilt angles of the sample; (c) incident electron energies.

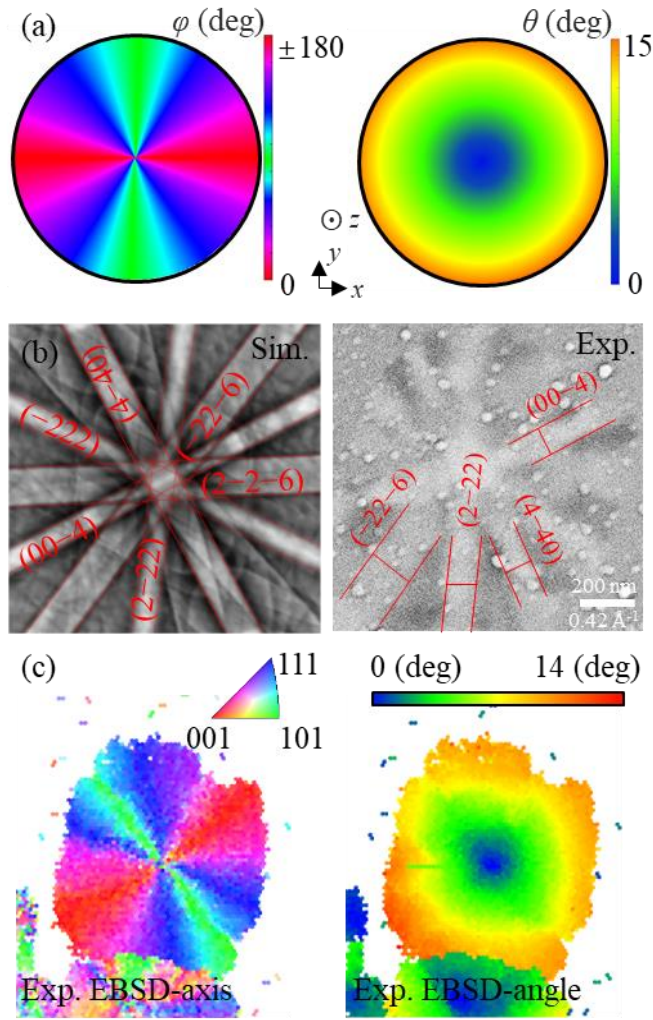

FIG. 3. (a) Crystal plane/orientation distribution diagram of an ideal CSRC island. The azimuthal angle $\varphi$ (left) and polar angle $\theta$ (right) of the crystal orientation at all positions on the crystal island are shown with the crystal orientation [$hkl$] at the center position as the $z$-axis. (b) Simulated Kikuchi pattern (left) and SEM image (right) obtained by BSE detector of a CSRC island with incident electron energy of 15 keV and crystal direction of [440] at the center position. (c) Axial (left) and angular (right) deviations in the grain reference orientation for EBSD measurements at the CSRC island in (b).

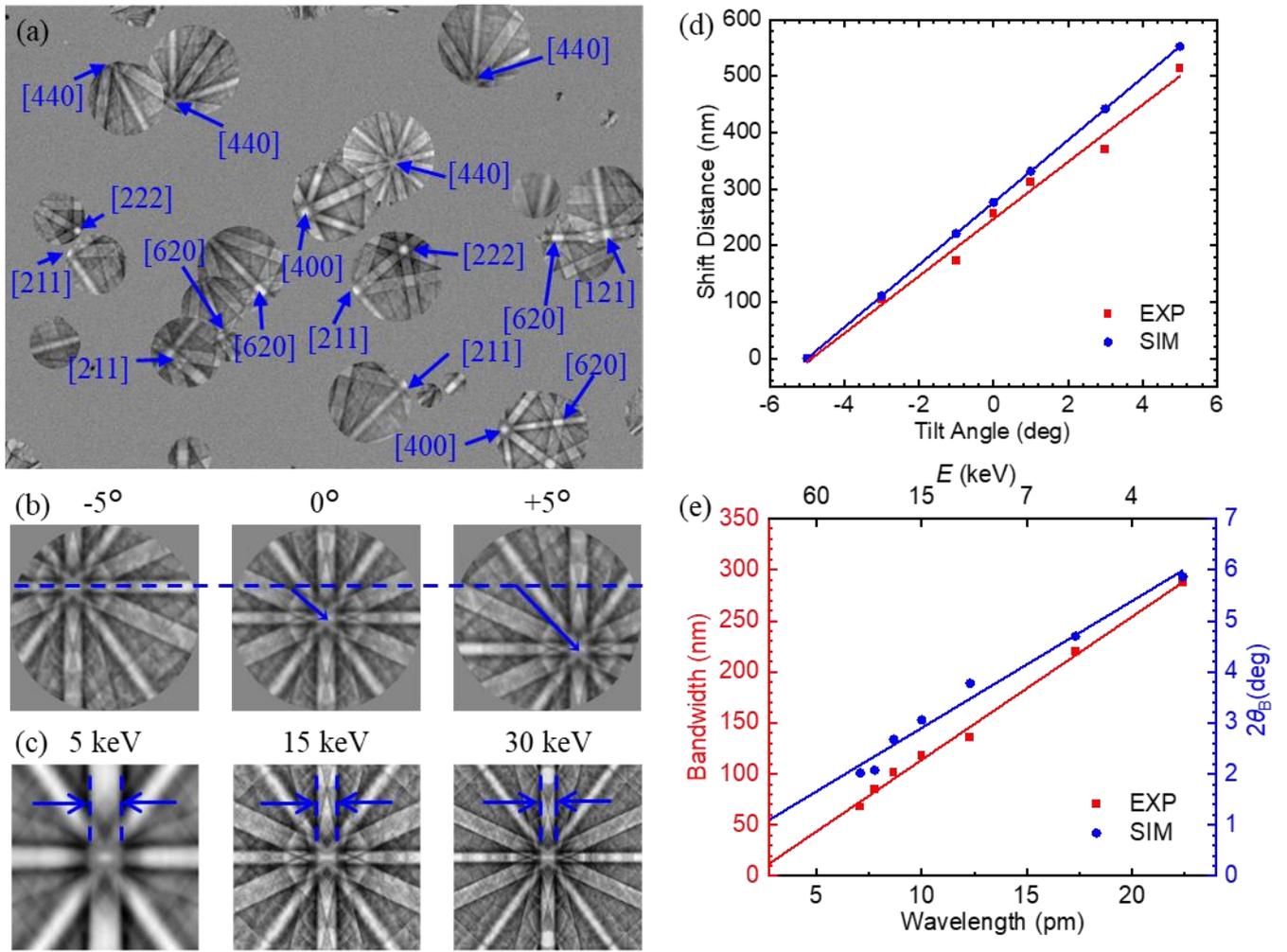

FIG. 4. (a) The image of the combined Kikuchi pattern of ideal CSRC islands simulated by the crystal orientation of each crystal island in Fig. 2(a). (b, c) Simulated Kikuchi patterns of [440] island with different (b) tilt angles of the sample; (c) incident electron energies. (d) Shift distance of the Kikuchi pattern versus tilt angle of CSRC island. (e) Variation of a selected bandwidth with incident electron energy/wavelength.